\documentclass[11pt,a4paper]{article}

\usepackage[utf8]{inputenc}
\usepackage[T1]{fontenc}
\usepackage{lmodern}
\usepackage{textcomp}
\usepackage{newunicodechar}
\newunicodechar{λ}{\lambda}
\DeclareSymbolFont{yhlargesymbols}{OMX}{yhex}{m}{n}
\DeclareMathAccent{\yhwidehat}{\mathord}{yhlargesymbols}{"62}

\usepackage[margin=2.5cm]{geometry}
\usepackage{graphicx}% Include figure files
\usepackage{dcolumn}% Align table columns on decimal point
\usepackage{bm}% bold math
\usepackage{amsmath,amsthm,amssymb}
\usepackage{wasysym} % for the diameter command
\usepackage{indentfirst} % indent first paragraph after \section
\usepackage[font=small,labelfont=bf]{caption}
\usepackage{authblk}
\usepackage[colorlinks=true,allcolors=blue]{hyperref}
\usepackage{doi}

\newcommand{\orcid}[1]{\,\href{https://orcid.org/#1}{\textsuperscript{\scriptsize ORCID}}}
\newcommand{\ack}[1]{\section*{Acknowledgments}#1}

\title{Finite-size scaling of spin-wave scattering and enhancement in magnonic Fabry--P\'erot nanoresonators}
\author{Anton Lutsenko\orcid{0009-0003-6536-7807}}
\author{Sebastiaan van Dijken\orcid{0000-0001-6372-2252}\thanks{Corresponding author: \href{mailto:sebastiaan.van.dijken@aalto.fi}{sebastiaan.van.dijken@aalto.fi}}}
\affil{NanoSpin, Department of Applied Physics, Aalto University School of Science, PO Box 15100, FI-00076 Aalto, Finland}
\date{}

\begin{document}

\maketitle

\begin{abstract}
Magnonic neural networks require compact resonant nodes that scatter spin waves efficiently, concentrate spin-wave energy and can be reconfigured. Hybrid magnonic Fabry–Pérot resonators, formed by a ferromagnetic-metal stripe dipolar-coupled to a yttrium iron garnet (YIG) film, provide low-loss and programmable control of spin-wave transmission, but have so far been studied only for stripes that are long compared with the spin-wave wavelength. Here we use micromagnetic simulations to determine how the response of a 200-nm-wide CoFeB stripe on a 55-nm-thick YIG film evolves as the stripe length is reduced from infinity to 640 nm. An infinitely long stripe acts as a one-dimensional cavity that opens a transmission gap at the first-order resonance (1.70 GHz and 1.47 GHz for the parallel and antiparallel magnetic configuration) and enhances the out-of-plane precession amplitude in the YIG film by factors of about 5 and 12, respectively, without distorting the transmitted wavefront. For lengths of a few micrometers, backward-volume spin waves emitted from the stripe ends modulate the precession along the stripe and fragment the reflected and transmitted wavefronts. Below about 1.2 µm, the resonator scatters incident spin waves over the entire isofrequency contour of the film, with caustic beams that are most pronounced in the antiparallel configuration. In this regime the parallel configuration supports several weakly enhanced quantized modes, whereas the antiparallel configuration supports a single mode with an out-of-plane enhancement of about 15. The width, thickness and shape of the stripe tune this mode by 50–100 MHz, and its strong enhancement lowers the onset of nonlinear frequency shift and fold-over to incident amplitudes of a few percent of the saturation magnetization. Sub-micrometer magnonic resonators are therefore promising reconfigurable nonlinear nodes for inverse-designed magnonic networks. 
\end{abstract}

\section{\label{sec:intro}Introduction}
Spin-wave resonators are essential building blocks of magnonic devices that can process information through wave scattering and interference \cite{Kruglyak2021,Chumak2022,Wang2024}. A resonator stores spin-wave energy in a small volume. This produces frequency-selective changes in the amplitude and phase of transmitted waves and raises the local precession amplitude, which in turn lowers the threshold for nonlinear processes such as amplitude-dependent frequency shifts and fold-over \cite{Wang2020,Fripp2023,Lutsenko2026}.

Several routes to spin-wave resonators have been explored. Resonators can be defined within the propagation medium itself, either by geometric confinement, as in YIG nanoring resonators side-coupled to a waveguide \cite{Wang2020}, by defects in magnonic crystals \cite{Baumgaertl2018}, or by local modifications of the internal magnetic field that partially reflect spin waves \cite{Kostylev2007}. Alternatively, a separately patterned magnetic element can be placed on top of a magnonic waveguide made of the same material and coupled to it through dynamic dipolar fields \cite{Kruglyak2021,Au2012a,Au2012b}. Such chiral magnonic resonators are operated near the ferromagnetic resonance of the magnetic element. Because the dipolar coupling between element and waveguide depends on the spin-wave propagation direction relative to the magnetization, they provide nonreciprocal control over spin-wave amplitude and phase. The resonant precession, however, resides largely in the magnetic element rather than in the waveguide \cite{Fripp2023}, which limits the precession enhancement within the waveguide.

The magnonic Fabry–Pérot resonator offers a different approach to spin-wave confinement \cite{Qin2021}. A ferromagnetic-metal stripe, made of CoFeB or permalloy, is patterned on a continuous YIG film and operated below its own ferromagnetic resonance frequency. Dynamic dipolar coupling between stripe and film lowers the spin-wave dispersion in the YIG beneath the stripe and makes it asymmetric with respect to the propagation direction. The two stripe edges therefore act as interfaces at which incident Damon–Eshbach (DE) spin waves of wavelength $\lambda_1$ convert into a shorter-wavelength mode ($\lambda_2$) or a much shorter-wavelength mode ($\lambda_3$), depending on the propagation direction. Spin waves circulating in the resonator acquire a round-trip phase ($|k_2| + |k_3|)w + \phi_0$, where $w$ is the stripe width and $\phi_0$ accounts for the phase acquired upon reflection. Destructive interference at a round-trip phase of (2$n$ + 1)$\pi$ opens narrow transmission gaps, whereas transmission at intermediate frequencies remains close to unity \cite{Qin2021}. Because $\lambda_3$ is much shorter than $\lambda_1$, a stripe only 200–300 nm wide controls spin waves with wavelengths of up to $\approx50$ µm. Moreover, the resonant energy is stored in the low-damping YIG rather than in the metal stripe \cite{Lutsenko2026}. Related YIG cavities bounded by permalloy barriers have been demonstrated experimentally \cite{Santos2023}, and the short-wavelength mode beneath metallic overlayers has been imaged directly \cite{Talapatra2023}. Magnonic Fabry–Pérot resonators have since been shown to act as non-volatile programmable $\pi$ phase shifters, switched by reversing the stripe magnetization relative to that of the film \cite{Lutsenko2025}, and as low-power nonlinear elements with neuron-like activation and frequency-selective nonlinear spin-wave absorption \cite{Lutsenko2026}.

These properties make magnonic resonators attractive as nodes of magnonic neural networks, in which spin-wave interference provides linear, all-to-all connectivity and resonators supply trainable weights and localized nonlinearity \cite{Kruglyak2021,Fripp2023,Papp2021,Breitbach2025}. Network parameters, including the positions and dimensions of individual resonators, can be optimized by inverse design \cite{Wang2021,Vilsmeier2026}. Such networks place three requirements on each node. Its footprint must be well below the spin-wave propagation length. It must scatter incident spin waves into a wide range of angles, so that each node couples to many others. Finally, it must strongly enhance the local precession, so that nonlinearity sets in at low incident power. A bistable magnetic state that switches the resonator response adds non-volatile programmability. Simulated networks of magnonic resonators based on 60-nm permalloy nanodisks on YIG satisfy the first two requirements \cite{Fripp2026a,Fripp2026b}, with nonlinearity provided by a low-frequency mode localized at the disk edges.

The magnonic Fabry–Pérot resonators studied so far do not meet the first requirement. Their stripes are 10–15 µm long \cite{Lutsenko2026,Lutsenko2025} or effectively infinite \cite{Qin2021}, so that they act as one-dimensional filters rather than localized scatterers. The evolution path of the physics of magnonic Fabry–Pérot resonators when the stripe length becomes smaller has not been established. A finite stripe length breaks the translational symmetry along the stripe and introduces static stray fields and edge regions, each of which can excite additional modes and redirect scattered spin waves.

Here we address this question by micromagnetic simulations of a 200-nm-wide, 20-nm-thick CoFeB stripe on a 55-nm-thick YIG film, with the stripe length reduced from infinity to 640 nm. We identify three regimes. For an infinitely long stripe, the resonator is an ideal one-dimensional Fabry–Pérot cavity that reflects and transmits incident spin waves without angular redistribution. For stripe lengths of a few micrometers, backward-volume (BV) spin waves emitted from the stripe ends modulate the resonant precession along the stripe and scatter spin waves predominantly along the stripe axis. Below about 1.2 µm, the resonator scatters spin waves over a wide range of angles, and its response depends qualitatively on the magnetic configuration. In the parallel configuration, the precession is quantized along the stripe, giving rise to several weak resonances whose frequencies increase as the stripe shortens. In the antiparallel configuration, a single strongly enhanced mode persists with only a moderate frequency shift. We attribute this difference to the static stray field of the stripe ends. We further show that the antiparallel nanoresonator can be tuned through its width, thickness and shape, and that its strong precession enhancement lowers the threshold for nonlinear effects.

\begin{figure}[h!]
	\centering
	\includegraphics[width=1\columnwidth]{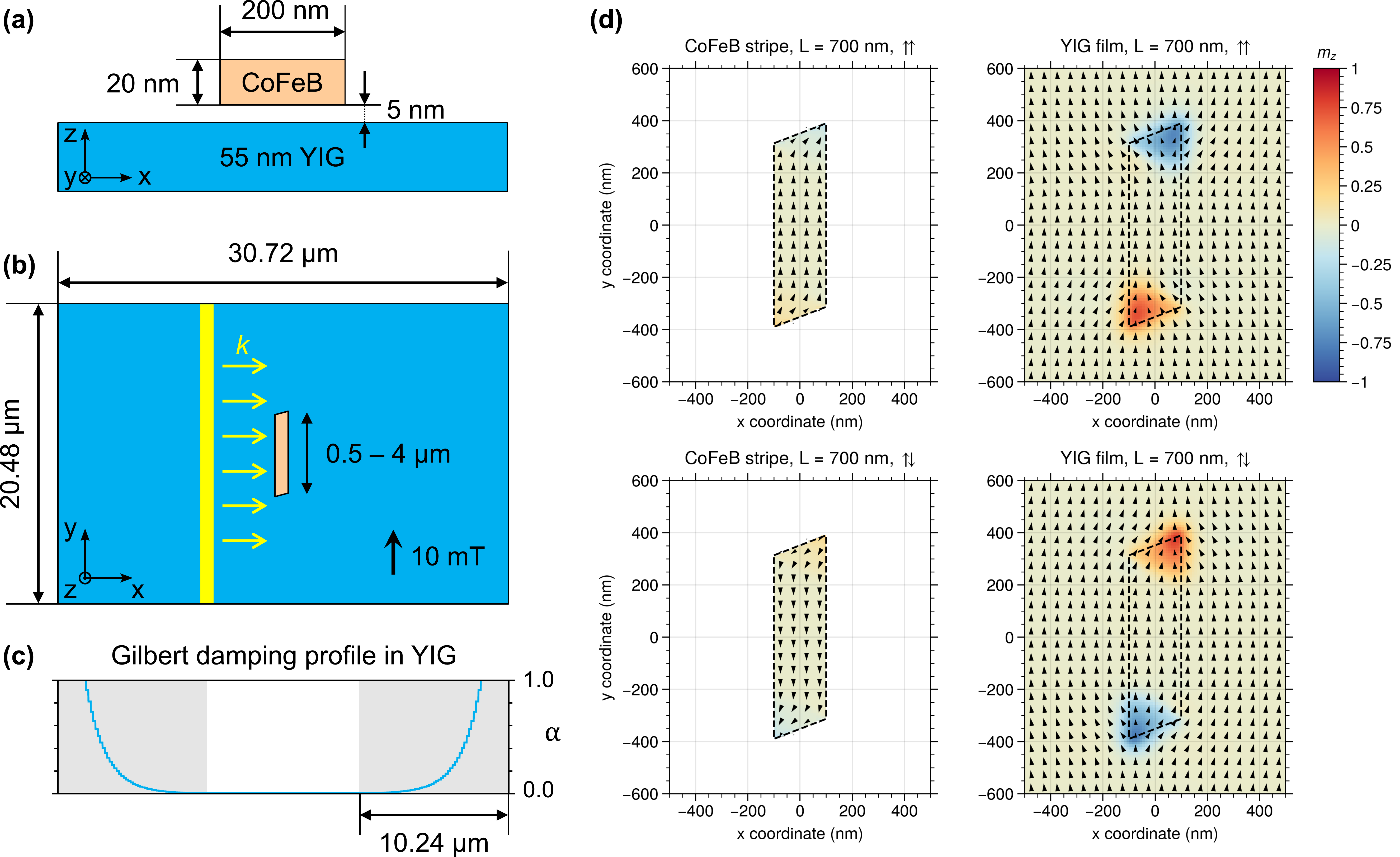}
	\caption{Magnonic Fabry–Pérot resonator and simulation geometry. (a) Cross-section of the resonator. A 20-nm-thick, 200-nm-wide CoFeB stripe is separated from a 55-nm-thick YIG film by a 5-nm nonmagnetic spacer. (b) Top view of the $30.72\times20.48$ µm$^2$ simulation area. A 500-nm-wide antenna (yellow), located 6 µm from the resonator, launches DE spin waves along $+x$. The stripe length $L$ along $y$ is varied between 640 nm and 4.68 µm, and a 10-mT bias field is applied along $+y$. (c) Profile of the Gilbert damping parameter along $x$. The damping increases exponentially within 10 µm ($x$) and 5 µm ($y$) of the boundaries to suppress spin-wave reflections. (d) Equilibrium magnetization of the bottom CoFeB layer (left) and top YIG layer (right) for $L=700$ nm in the P (top) and AP (bottom) configurations. Arrows indicate the in-plane magnetization and the color scale the out-of-plane component $m_z$. The stray field of the CoFeB stripe ends tilts the YIG magnetization out of the film plane, with opposite sign for the two configurations.}
	\label{Fig1-crossection}
\end{figure}

\section{\label{sec:methods}Model and methods}
\textit{Geometry.} The simulated resonator consists of a 20-nm-thick, 200-nm-wide CoFeB stripe separated from a 55-nm-thick YIG film by a 5-nm nonmagnetic spacer (Fig. \ref{Fig1-crossection}(a)), which suppresses interlayer exchange coupling while transmitting dynamic dipolar fields \cite{Lutsenko2026,Qin2021,Lutsenko2025}. We denote by $w$ the stripe width along $x$, the propagation direction of the incident spin waves, and by $L$ the stripe length along $y$, the direction of the 10-mT bias field (Fig. \ref{Fig1-crossection}(b)). $L$ is varied between 640 nm and 4.68 µm and compared with an infinitely long stripe. The short edges of the stripe are slanted by $\pi$/8 with respect to $x$, which lifts the degeneracy between end-domain configurations and yields a reproducible magnetic ground state. The bias field saturates the YIG film and establishes the  DE geometry ($k\perp M$) for incident spin waves. At 10 mT the ferromagnetic resonance of an extended CoFeB film lies at about 3.5 GHz \cite{Kittel1948}, and the shape anisotropy of the stripe raises it further. The stripe is therefore driven well below its own resonance in the 1.2–2.2 GHz range studied here, as required for Fabry–Pérot operation \cite{Qin2021}.

\textit{Micromagnetics.} Simulations were performed with MuMax3 \cite{Vansteenkiste2014}. The simulation area of $30.72\times20.48$ µm$^2$ was discretized into $3072\times2048\times16$ cells of $10\times10\times5$ nm$^3$, comprising 11 layers of YIG, one spacer layer, and four layers of CoFeB. Periodic boundary conditions were applied in the $x$ and $y$ directions, and spin-wave reflections from the boundaries were suppressed by a Gilbert damping parameter that increases exponentially over 10 µm in $x$ and 5 µm in $y$ toward the edges of the simulation area (Fig. \ref{Fig1-crossection}(c)). For YIG, we used a saturation magnetization $M_{s}=1.4\times10^5$ A/m, an exchange stiffness $A_{ex}=3.5\times10^{-12}$ J/m \cite{Klingler2015} and a Gilbert damping parameter $\alpha=2\times10^{-4}$ \cite{Onbasli2014}. For CoFeB, we used $M_{s}=1.15\times10^6$ A/m, $A_{ex}=1.6\times10^{-11}$ J/m and $\alpha=5\times10^{-3}$ \cite{Qin2021,Lutsenko2025}. The YIG magnetization was aligned with the bias field along $+y$, and the stripe magnetization was initialized along $+y$ (parallel (P) configuration) or $-y$ (antiparallel (AP) configuration). The system was relaxed to its ground state before excitation. In both configurations the stray field of the magnetic charges at the stripe ends tilts the YIG magnetization out of the film plane beneath the ends, with opposite sign for the two configurations (Fig. \ref{Fig1-crossection}(d)). To separate static and dynamic contributions of the CoFeB stripe, additional simulations were performed in which the CoFeB magnetization was frozen after relaxation using the \texttt{FrozenSpins} option of MuMax3). 

\textit{Spin-wave excitation.} Spin waves were excited by the Oersted field of an infinitely long Au antenna, 500 nm wide and 120 nm thick, placed 5 nm above the YIG film and 6 µm from the resonator. The field was calculated from the Biot–Savart law for a uniform current density and truncated at 1 µm from the antenna to avoid direct excitation of the resonator. In Sections \ref{sec:width-height} and \ref{sec:ellipse}, a spatially uniform in-plane field along $x$ confined to the CoFeB stripe was applied instead, which isolates the intrinsic resonator response from its coupling to incident spin waves. For broadband spectra, the excitation followed a sinc function with a cutoff frequency of 4 GHz centered 4 ns after the start of the simulation, and the dynamic magnetization was recorded for 64 ns at a sampling rate of 16 GHz, which gives a frequency resolution of 15.6 MHz. Complex amplitudes at each frequency were obtained by a discrete Fourier transform along the time axis. For the nonlinear simulations in Section \ref{sec:nonlin}, a continuous sinusoidal excitation was applied at fixed frequency.

\begin{figure}
	\centering
	\includegraphics[width=1.0\columnwidth]{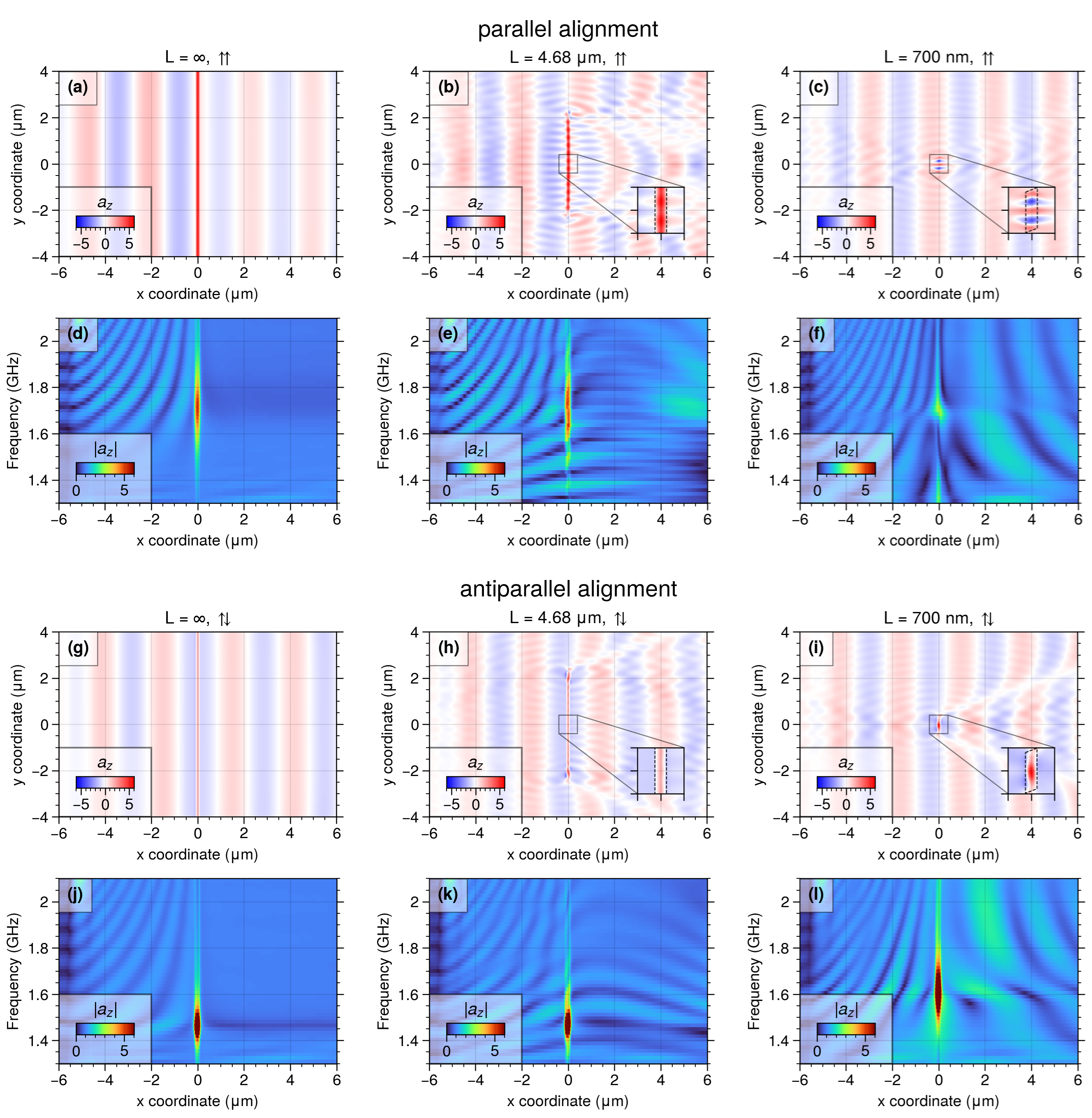}
	\caption{Spin-wave scattering and magnetization precession enhancement for three resonator lengths. (a–c) Real part of $a_z$ in the top YIG layer at 1.70 GHz for the P configuration and $L=\infty$ (a), $L=4.68$ µm (b) and $L=700$ nm (c). Insets show the resonator region at higher magnification. (d–f) Normalized amplitude $|a_z|$ along $y=0$ as a function of frequency for the same resonators. The stripe is located at $x=0$ and spin waves propagate from left to right. Fringes at $x<0$ originate from the interference of incident and reflected spin waves, and the dark band at $x>0$ in (d) marks the $n=1$ Fabry–Pérot transmission gap. (g–l) Same as (a-f) for the AP configuration. For $L=\infty$ and $L=4.68$ µm the resonance shifts to lower frequency and the precession enhancement increases compared with the P configuration. For $L=700$ nm the strongest enhancement occurs at a higher frequency, and the scattered spin waves form oblique caustic beams.}
	\label{Fig2_len_comparison}
\end{figure}

\textit{Analysis.} The spin-wave and precession amplitudes in Figs. \ref{Fig2_len_comparison} to \ref{all-lengths-spectra-comparison-par} and \ref{frozen-spins} were normalized to a reference simulation without the CoFeB stripe, performed with identical excitation at the same frequency. The reference amplitude was taken at the center of the top YIG layer, $\mathbf r_0$, which corresponds to the center of the resonator in the simulations with the stripe. The normalized complex amplitudes are defined as

\begin{equation}
a_{x,z}(\mathbf r, f) = \frac{m_{x,z}(\mathbf r, f)}{\left|m_{x,z}^{\mathrm{ref}}(\mathbf r_0, f)\right|},
\end{equation}

\noindent so that $|a_{x,z}|$ maps the local spin-wave amplitude relative to that of the incident spin wave, and the real part of $a_{x,z}$ visualizes the instantaneous spin-wave pattern. Because the dipolar coupling to the stripe strongly changes the precession ellipticity \cite{Qin2021}, the magnetization components along $x$ and $z$ are reported separately. For the uniform excitation of the CoFeB stripe in Figs. \ref{varying-widths-heights-comparison} and \ref{ellipse-widths-heights-comparison}, no incident spin wave exists, and the precession spectra are therefore given as Fourier amplitudes in arbitrary units. In the nonlinear simulations of Fig. \ref{nonlinear}, the precession amplitudes are given as $|m_{x,z}| = |M_{x,z}|/M_s$, i.e., normalized to the saturation magnetization of YIG. Isofrequency contours of the scattered spin waves were obtained from two-dimensional spatial Fourier transforms of $a_z$ in the top YIG layer. Dispersion relations of the continuous CoFeB/YIG bilayer were calculated with the finite-element package TetraX \cite{Korber2022a,Korber2022b}.

\section{\label{sec:general}From an infinite to a finite magnonic Fabry–Pérot resonator}
Shortening the CoFeB stripe turns a one-dimensional magnonic Fabry–Pérot resonator into a two-dimensional scatterer. Figure \ref{Fig2_len_comparison} compares the real part of $a_z$ for an infinitely long stripe, a 4.68-µm-long stripe and a 700-nm-long stripe at 1.70 GHz, the frequency of the $n=1$ Fabry–Pérot resonance of the infinite stripe in the P configuration, together with the frequency-resolved amplitude $|a_z|$ along $y=0$, for the P (Fig. \ref{Fig2_len_comparison}(a)-(f)) and AP (Fig. \ref{Fig2_len_comparison}(g)-(l)) configurations. Figures \ref{Fig3_slice_comparison_4um} and \ref{Fig4_freq_comparison_600nm} resolve the few-micrometer and sub-micrometer cases in more detail. For each regime we first describe the spin precession within the resonator, that is, in the YIG film beneath the CoFeB stripe, and then the pattern of reflected, transmitted and scattered spin waves in the surrounding YIG film.

\subsection{\label{subsec:length-infinite}Infinitely long stripes: a one-dimensional Fabry–Pérot cavity}
For $L=\infty$ (Fig. \ref{Fig2_len_comparison}(a),(d),(g),(j)) the structure approximates the magnonic Fabry–Pérot resonator studied in Ref. \cite{Qin2021}. Translational symmetry along $y$ makes the resonance uniform along the stripe. Its only spatial structure is across the 200-nm width, where the counterpropagating $λ_2$ and $λ_3$ waves interfere. For $w=200$ nm the resonance at 1.70 GHz in the P configuration is of first order ($n=1$), in agreement with the $n=1$ transmission gaps measured for 200–300-nm-wide CoFeB stripes \cite{Qin2021}. At resonance, $|a_z|$ beneath the stripe reaches about 5, which shows that the cavity accumulates spin-wave energy in the YIG film and enhances the magnetization precession well above the amplitude of the incident spin wave. In the AP configuration the resonance shifts down to 1.47 GHz, becomes narrower and its enhancement more than doubles, in agreement with the behavior of long resonators \cite{Lutsenko2026,Qin2021,Lutsenko2025}. The downshift follows from the larger magnitude of $|k_3|$ at a given frequency in the AP configuration, which satisfies the resonance condition at a lower frequency \cite{Qin2021}.

Outside the resonator, the conservation of $k_y$ restricts the scattering of the incident DE spin wave to reflection ($k_x = -k_1$) and transmission ($k_x = +k_1$). In front of the stripe, the reflected and incident spin waves form a standing-wave pattern, which appears as fringes for $x < 0$ in Fig. \ref{Fig2_len_comparison}(d),(j) whose contrast is largest near resonance. Behind the stripe, the transmitted wavefront remains flat, and its amplitude is reduced in a narrow band around the resonance, which constitutes the Fabry–Pérot transmission gap (dark band for $x > 0$ in Fig. 2(d),(j)). An infinitely long resonator acts as a frequency filter and phase shifter for an incident plane wave \cite{Qin2021,Lutsenko2025} without redistributing spin-wave energy into other directions.

\subsection{\label{subsec:length-long}Few-micrometer-long stripes: Fabry–Pérot resonance dressed by backward-volume waves}
\begin{figure}
	\centering
	\includegraphics[width=\columnwidth]{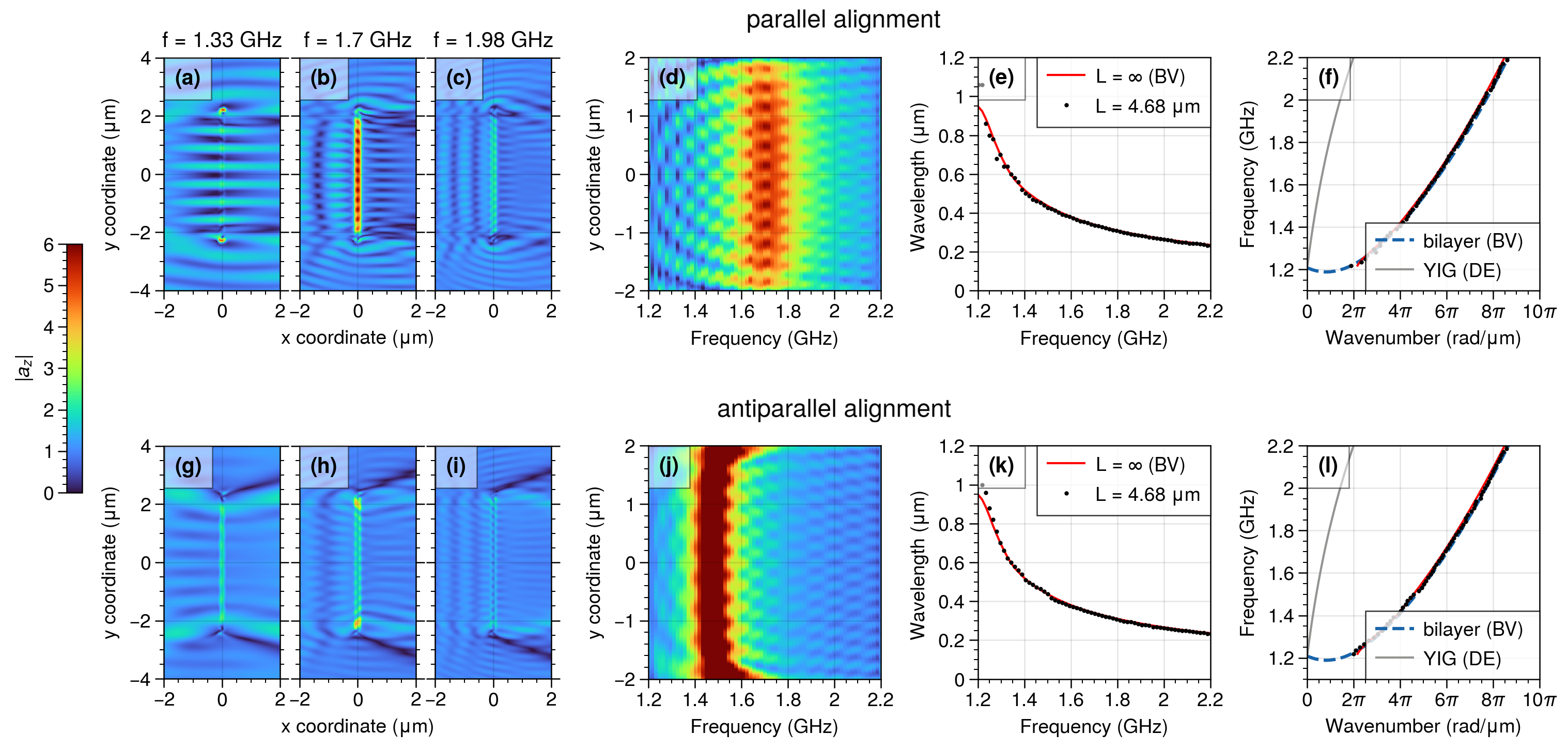}
	\caption{Backward-volume modulation of the Fabry–Pérot resonance in a few-micrometer-long resonator. (a-c) Normalized amplitude $|a_z|$ in the top YIG layer for $L=4.68$ µm and the P configuration at 1.33 GHz (a), 1.70 GHz (b), and 1.98 GHz (c). (d) Normalized amplitude $|a_z|$ along $x=0$ within the resonator as a function of frequency, showing the modulation of the magnetization precession along $y$. (e) Spin-wave wavelength extracted from the modulation and (f) corresponding wavenumber (black symbols), compared with the BV dispersion obtained from a micromagnetic simulation of an infinitely long resonator (red line) and from a TetraX calculation for a continuous YIG/CoFeB bilayer (dashed blue line). The agreement identifies BV spin waves emitted from the stripe ends as the origin of the modulation. The DE dispersion of the bare YIG film (gray line) is shown as reference. (g–l) Same as (a-f) for the AP configuration.} 
	\label{Fig3_slice_comparison_4um}
\end{figure}

For $L$ of a few micrometers, the Fabry–Pérot resonance persists at nearly the same frequency, but the magnetization precession beneath the stripe is modulated along $y$ (insets of Fig. \ref{Fig2_len_comparison}(b),(h)). This modulation is present at all frequencies (Fig. \ref{Fig3_slice_comparison_4um}(a)–(c)), and its period decreases with increasing frequency (Fig. \ref{Fig3_slice_comparison_4um}(d)). It cannot be attributed to standing modes of the CoFeB stripe itself, whose frequencies lie above the stripe ferromagnetic resonance and thus well outside the frequency range studied. We attribute the modulation to the interference between the Fabry–Pérot resonance, which is uniform along $y$, and BV spin waves emitted at the stripe ends that propagate along the resonator. The stripe ends act as sources of BV spin waves because both the dynamic dipolar coupling and the static stray field change abruptly there, in analogy to spin-wave emission from regions of nonuniform internal field \cite{Kostylev2007,Davies2015,Mushenok2017}. 
The spin-wave wavelength extracted from the modulation period at each frequency (Fig. \ref{Fig3_slice_comparison_4um}(e)) and the corresponding dispersion (Fig. \ref{Fig3_slice_comparison_4um}(f)) indeed coincide with the BV dispersion obtained from a micromagnetic simulation of an infinitely long resonator and from a dynamic-matrix calculation for a continuous CoFeB/YIG bilayer. The agreement holds for both magnetic configurations (AP data in Fig. \ref{Fig3_slice_comparison_4um}(g)–(l)). Moreover, the BV dispersion of the bilayer is almost identical to that of the bare YIG film. This is expected because the dynamic dipolar coupling between stripe and film is chiral, being strongest for $k \perp M$ and vanishing for $k \parallel M$ \cite{Kruglyak2021}. BV spin waves therefore propagate along the resonator nearly unaffected by the CoFeB, except in the vicinity of the stripe ends.

Because the finite stripe breaks the translational symmetry along $y$, the resonator now scatters incident spin waves into a range of $k_y$. Three features are apparent in Figs. \ref{Fig2_len_comparison} and \ref{Fig3_slice_comparison_4um}. The reflected and transmitted spin waves inherit the modulation along $y$, so that the wavefronts on both sides of the stripe are fragmented rather than flat. The Fabry–Pérot transmission gap is less pronounced than for $L=\infty$, because the attenuated transmitted spin wave behind the stripe interferes with wave modulations along $y$ as well as with diffracted waves from the stripe ends. Finally, spin waves propagate from the stripe ends into the YIG film, predominantly along $\pm y$. As these spin waves travel nearly parallel to the magnetization, they are of BV character. A resonator of a few micrometers length therefore converts part of the incident DE spin wave into BV spin waves. These features are observed for both magnetic configurations.

\begin{figure}
	\centering
	\includegraphics[width=1\columnwidth]{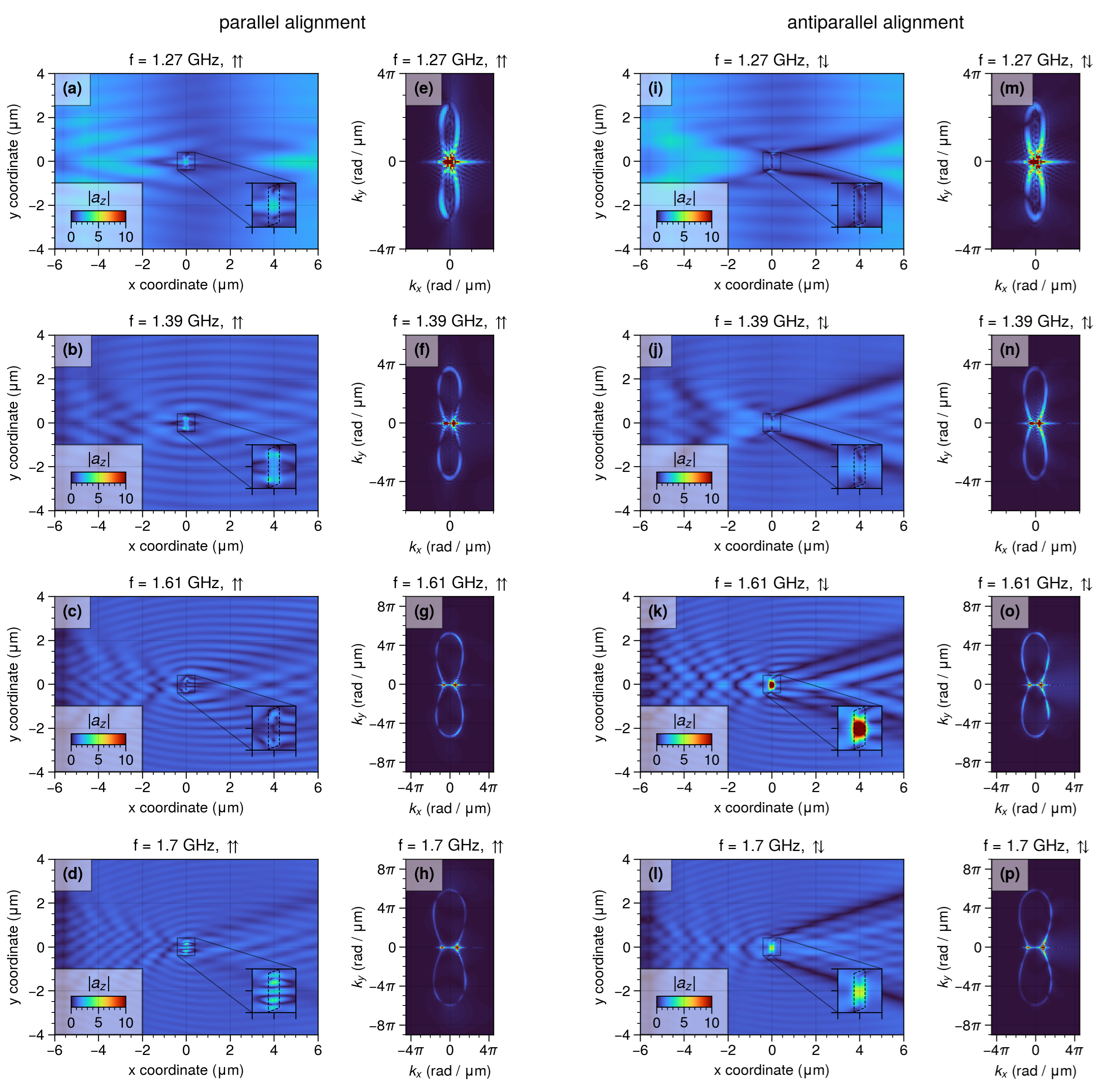}
	\caption{Quantized modes, a strongly enhanced single mode and wide-angle spin-wave scattering in a 700-nm-long resonator. (a-d) Normalized amplitude $|a_z|$ in the top YIG layer for the P configuration at 1.27 GHz (a), 1.39 GHz (b), 1.61 GHz (c), and 1.70 GHz (d). Insets show $|a_z|$ within the resonator, revealing modes quantized along $y$. (e–h) Magnitude of the spatial Fourier transform of $a_z$ at the same frequencies. The scattered intensity is distributed along the isofrequency contour of the YIG film. (i–p) Same as (a-h) for the AP configuration, in which a single mode without nodes along $y$ is excited at 1.61 GHz (k) and the scattered spin waves form caustic beams (i–l).}
	\label{Fig4_freq_comparison_600nm}
\end{figure}

\subsection{\label{subsec:length-short}Sub-micrometer stripes: quantized modes, a strongly enhanced single mode and wide-angle spin-wave scattering}
Below $L \approx 1.2$ µm, the response of the resonator depends qualitatively on the magnetic configuration (Figs. \ref{Fig2_len_comparison}(c),(f),(i),(l) and \ref{Fig4_freq_comparison_600nm}). In the P configuration, the Fabry–Pérot resonance of the infinite stripe is no longer observed. 

Instead, two sharp resonances appear at 1.27 and 1.70~GHz, whose precession profiles are quantized along $y$ (insets of Fig. \ref{Fig4_freq_comparison_600nm}(a),(d)), together with a broad, weaker feature with a maximum near 1.39~GHz, where the precession is concentrated near the slanted stripe ends (inset of Fig. \ref{Fig4_freq_comparison_600nm}(b)). Away from these resonances the resonator is only weakly excited (inset of Fig. \ref{Fig4_freq_comparison_600nm}(c)). In the AP configuration, a single mode at 1.61 GHz dominates (Fig \ref{Fig4_freq_comparison_600nm}(k)). Its precession is maximal at the stripe center, it has no nodes along $y$, and its amplitude exceeds that of all modes in the P configuration. Compared with the infinite stripe, the AP resonance has moved up from 1.47 to 1.61 GHz (Fig. \ref{Fig2_len_comparison}(j),(l)). The origin of this shift and of the difference between the two magnetic configurations is discussed in Section \ref{sec:length}.

A sub-micrometer stripe acts as a localized scatterer. Because its length is shorter than the incident spin-wave wavelength, the resonator couples to all wave vectors allowed at the excitation frequency, and the amplitude maps show scattered spin waves spreading over a wide angular range in front of and behind the resonator (Fig. \ref{Fig4_freq_comparison_600nm}(a)–(d),(i)–(l)). Spatial Fourier transforms of $a_z$ confirm this behavior (Fig. \ref{Fig4_freq_comparison_600nm}(e)–(h),(m)–(p)). The scattered intensity is distributed along the entire isofrequency contour of the YIG film and thus comprises DE-like ($k_x$-dominated), BV-like ($k_y$-dominated) and oblique components. The contour has the shape of a figure eight elongated along $k_y$, reflecting the anisotropic dipolar dispersion of an in-plane magnetized film \cite{Damon1961,Kalinikos1986}. Where the isofrequency contour has inflection points, the group velocities of many wave vectors converge, and scattered energy is concentrated in caustic beams \cite{Wang2024,Lutsenko2025,Veerakumar2006,Schneider2010,Gieniusz2013}. These beams are clearly resolved as narrow oblique rays in the AP configuration (Figs. \ref{Fig2_len_comparison}(i) and \ref{Fig4_freq_comparison_600nm}(i)–(l)). In this configuration the total scattered intensity is equal to or larger than in the P configuration.

The scattered intensity along the isofrequency contour is not uniform but exhibits minima whose number and positions depend on frequency, magnetic configuration and stripe length. These minima are asymmetric with respect to both $k_x$ and $k_y$. The asymmetry in $k_x$ is a direct consequence of the chirality of the dipolar coupling between the stripe and film, which also produces the nonreciprocal Fabry–Pérot response \cite{Qin2021}. An asymmetry in $k_y$, by contrast, is not expected for a rectangular stripe, because for $M \parallel y$ the system is invariant under the mirror operation $y \rightarrow -y$. The slanted stripe ends break this symmetry, and we therefore attribute the asymmetry in $k_y$ primarily to the stripe shape. The angular distribution of scattered spin waves can thus be engineered by the shape of the stripe ends, which provides an additional degree of freedom for the inverse design of magnonic networks.

\section{\label{sec:enhancement}Magnetization precession enhancement within the resonator}
\begin{figure}[h]
	\centering
	\includegraphics[width=\columnwidth]{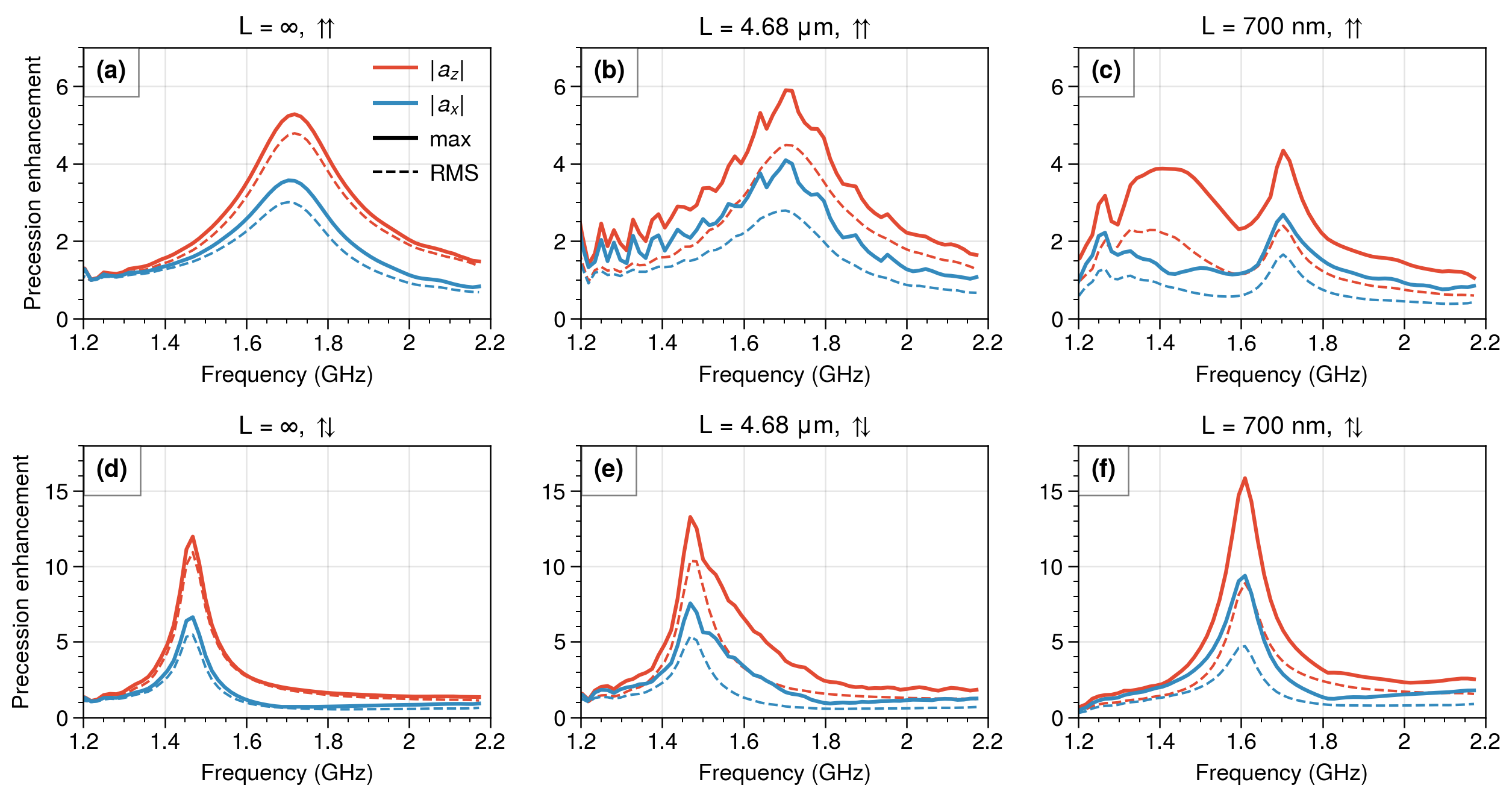}
	\caption{Magnetization precession enhancement spectra for three resonator lengths. Maximum (solid lines) and root-mean-square (dashed lines) values of the normalized amplitudes $|a_z|$ (red) and $|a_x|$ (blue) in the top YIG layer beneath the CoFeB stripe for $L=\infty$ (a,d), $L=4.68$ µm (b,e) and $L=700$ nm (c,f), in the P (a–c) and AP (d–f) configurations. Note the different vertical scales for the two configurations.}
	\label{Fig5_three-spectra-comparison}
\end{figure}

For an infinitely long resonator, the complex transmission coefficient fully characterizes the response \cite{Qin2021}. For a finite resonator, transmission is no longer a well-defined single quantity, because the scattered amplitude depends on the observation point. We therefore characterize the resonators by the enhancement of magnetization precession within them, evaluated as the maximum and root-mean-square values of $|a_x|$ and $|a_z|$ in the top YIG layer beneath the stripe. This quantity can be compared across all lengths and, as shown below, determines the threshold for nonlinear effects. Figure \ref{Fig5_three-spectra-comparison} presents enhancement spectra for three representative lengths.

In the P configuration, the infinite resonator exhibits a single, nearly Lorentzian peak at 1.70 GHz with a maximum $|a_z|$ of about 5.2 (Fig. \ref{Fig5_three-spectra-comparison}(a)). The maximum and root-mean-square values almost coincide because the precession is uniform along $y$. For $L=4.68$ µm the BV modulation superimposes oscillations on the maximum enhancement (Fig. \ref{Fig5_three-spectra-comparison}(b)), whereas the root-mean-square enhancement remains close to that of the infinite resonator. The BV spin waves thus redistribute the precession along the stripe rather than adding energy to the resonator. 
For $L = 700$ nm the Fabry–Pérot resonance is replaced by two quantized modes at 1.27 and 1.70 GHz and a broad feature near 1.39 GHz (Fig. \ref{Fig5_three-spectra-comparison}(c)), all of which are enhanced less than the resonance of the infinite stripe. The feature at 1.39 GHz appears predominantly in $|a_z|$, whereas the quantized modes are represented in both components. Because the precession of the 1.39-GHz mode is concentrated near the resonator ends, where the static stray field from the CoFeB stripe tilts the YIG magnetization out of the film plane (Fig.  \ref{Fig1-crossection}(d)), its precession cone is tilted with respect to that of the incident spin wave, which increases the projection onto the $z$ axis.

In the AP configuration the trend is reversed. The infinite resonator shows a narrower and stronger peak at 1.47 GHz, with a maximum $|a_z|$ of about 12 (Fig. \ref{Fig5_three-spectra-comparison}(d)). For $L=4.68$ µm the peak remains at nearly the same frequency, acquires a tail on its high-frequency side and is two to three times stronger than in the P configuration, while the oscillations caused by BV spin waves are weaker (Fig. \ref{Fig5_three-spectra-comparison}(e)). For $L=700$ nm a single resonance at 1.61 GHz dominates the spectrum, and its maximum $|a_z|$ of about 15 and maximum $|a_x|$ of about 10 exceed the enhancement of the infinite resonator (Fig. \ref{Fig5_three-spectra-comparison}(f)). Shortening the stripe in the AP configuration therefore increases, rather than reduces, the precession enhancement within the resonator.

\section{\label{sec:length}Dependence of precession enhancement on resonator length}
\begin{figure}[h]
	\centering
	\includegraphics[width=\columnwidth]{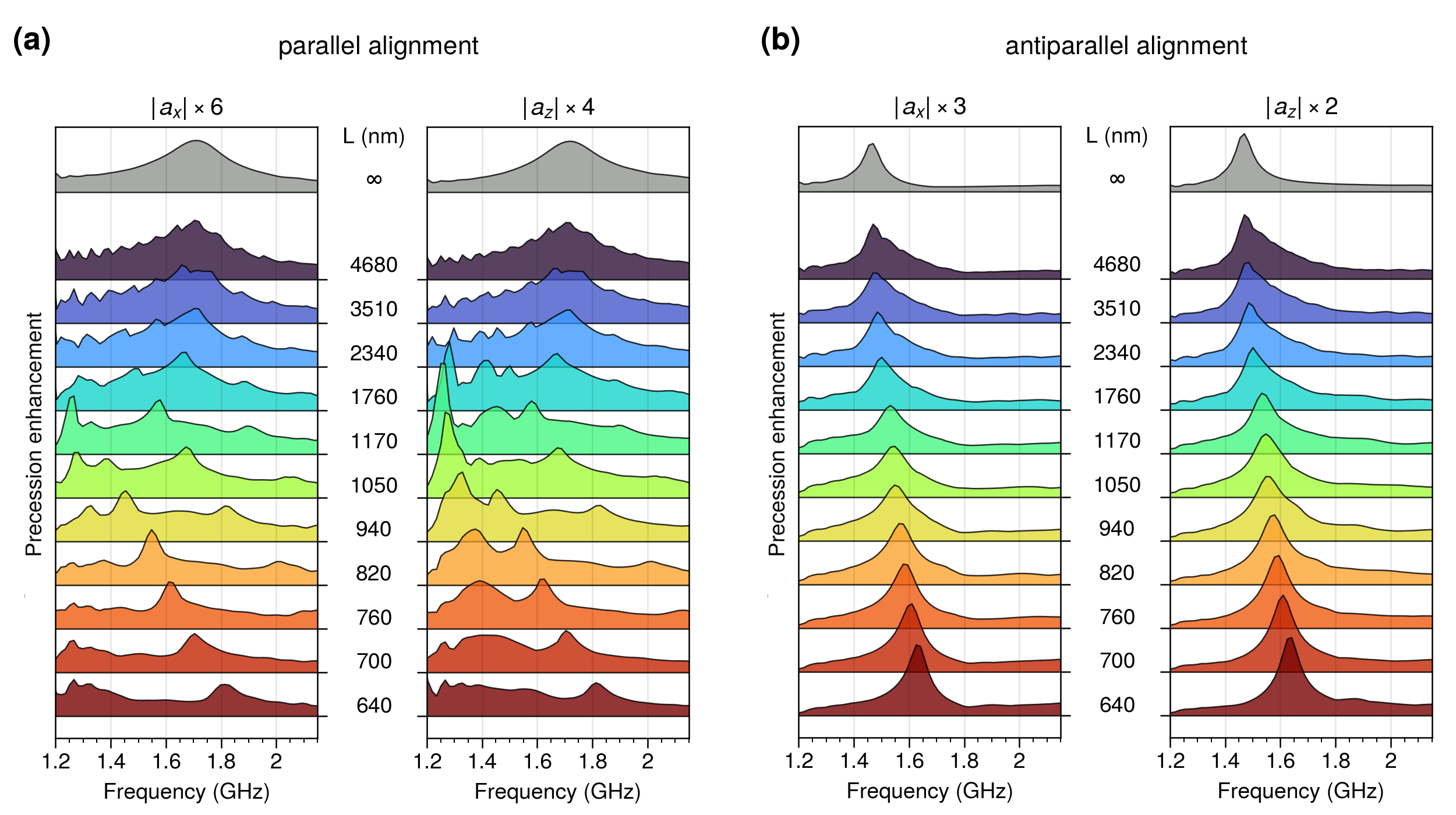}
	\caption{Dependence of the magnetization precession enhancement on resonator length. Maximum $|a_x|$ (left) and $|a_z|$ (right) as a function of frequency for $L$ between 640 nm and 4.68 µm in the P (a) and AP (b) configurations, with spectra for $L=\infty$ shown in gray. Spectra are multiplied by the indicated factors and vertically offset for clarity. In the P configuration the Fabry–Pérot resonance breaks up into quantized modes below $L\approx1.2$ µm, whereas in the AP configuration a single resonance persists, shifts to higher frequency and gains amplitude as $L$ decreases.}
	\label{all-lengths-spectra-comparison-par}
\end{figure}

Figure \ref{all-lengths-spectra-comparison-par} shows magnetization precession enhancement spectra for eleven stripe lengths between 4.68 µm and 640 nm. In both magnetic configurations a crossover occurs near $L\approx1.2$ µm, but its character differs fundamentally. In the P configuration (Fig. \ref{all-lengths-spectra-comparison-par}(a)), the spectra for $L=$ 4.68 to 1.76 µm are dominated by the $n=1$ Fabry–Pérot resonance near 1.70 GHz. The superimposed oscillations caused by BV spin waves become sparser with decreasing $L$, because fewer BV wavelengths fit along the stripe. Between 1.76 and 1.17 µm the Fabry–Pérot resonance collapses, and for shorter stripes the spectrum consists of discrete resonances whose frequencies increase rapidly with decreasing $L$, as expected for modes with $k_y \propto m/L$, where $m$ is the mode index. The strongest enhancement of the entire series occurs for $L=1.17$ µm at the low-frequency end of the spectrum, and the enhancement decreases monotonically for shorter stripes. In the AP configuration (Fig. \ref{all-lengths-spectra-comparison-par}(b)), the main resonance of the infinite stripe persists at all lengths. Below $L \approx 1.2$ µm it shifts gradually from 1.47 to 1.63 GHz, remains the single dominant resonance and grows in amplitude, whereas higher-order modes remain weak.

Two length-dependent effects act on the YIG beneath the stripe. The first is the quantization of the precession along $y$. Once $L$ becomes smaller than about 20 BV wavelengths ($\approx 0.33$ µm at 1.70 GHz; Fig. \ref{Fig3_slice_comparison_4um}(e)), the Fabry–Pérot precession acquires discrete $k_y$ components, and confinement along $y$ combines with Fabry–Pérot interference along $x$. Quantization alone, however, does not explain why the two magnetic configurations respond so differently below $L \approx 1.2$ µm. The second effect, which distinguishes them, is the static stray field of the stripe. An infinitely long stripe magnetized along $y$ carries no magnetic charges and produces no stray field in the YIG film. A finite stripe carries magnetic charges of density $\pm M_{s,CoFeB}$ on its ends, whose field beneath the stripe is directed opposite to the stripe magnetization. Treating these charges as point charges gives $\mu_0 H_s \approx 2 \mu_0 M_{s,CoFeB} w h/(\pi L^2)$ at the stripe center, where $h$ is the stripe thickness, which predicts a stray field that grows rapidly as the stripe shortens.

\begin{figure}[h]
	\centering
	\includegraphics[width=\columnwidth]{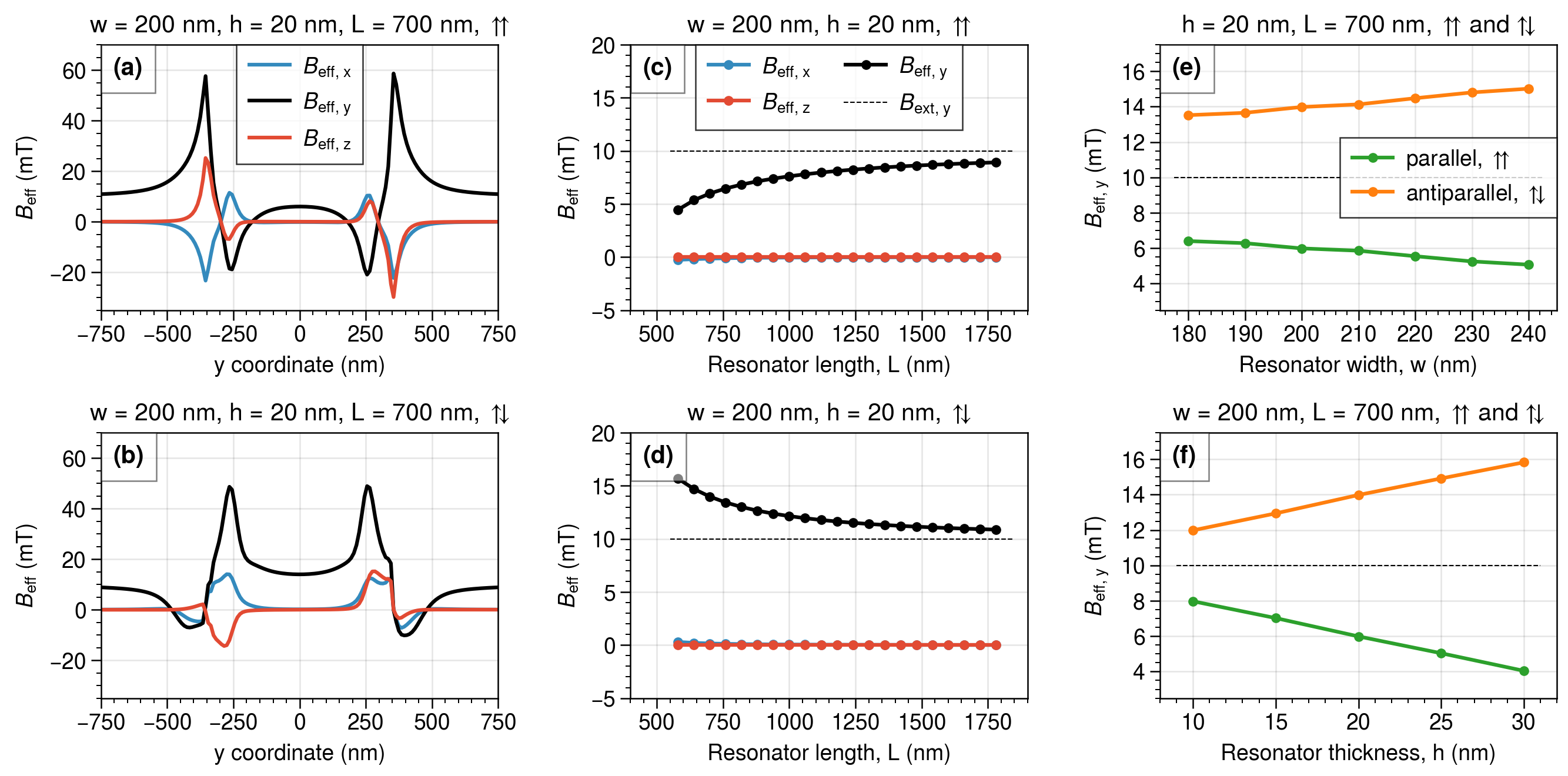}
	\caption{Static effective magnetic field in the YIG film. (a,b) Components $B_\textrm{eff,x}$ (blue), $B_\textrm{eff,y}$ (black) and $B_\textrm{eff,z}$ (red) of the static effective field in the top YIG layer along the stripe axis ($x=0$) for a 700-nm-long stripe ($w=200$ nm, $h=20$ nm) in the P (a) and AP (b) configurations. (c,d) Effective field components in YIG at the stripe center ($x=y=0$) as a function of stripe length $L$ for $w=200$ nm and $h=20$ nm in the P (c) and AP (d) configurations. The dotted line marks the 10-mT bias field. (e,f) $y$ component of the effective field in YIG at the stripe center for $L=700$ nm as a function of stripe width $w$ at $h=20$ nm (e) and of stripe thickness $h$ at $w=200$ nm (f), for both configurations. The stray field of the stripe reduces the field beneath the stripe in the P configuration and increases it in the AP configuration, and its effect grows as $L$ decreases and as $w$ and $h$ increase.}
	\label{effective-field-comparison}
\end{figure}

Figure \ref{effective-field-comparison} shows the static effective field in the top YIG layer obtained from the relaxed magnetic configuration. Along the stripe axis ($x=0$), the field profiles of the two configurations are qualitatively different (Fig. \ref{effective-field-comparison}(a),(b)). In the P configuration, the $y$ component of the effective field beneath the stripe is reduced from the 10-mT bias field to about 6 mT at the stripe center and becomes negative, down to about $-20$ mT, just inside the stripe ends. Just outside the ends, the field of the end charges adds to the bias field and produces narrow maxima of about 60 mT. The YIG beneath the stripe thus experiences a field well that is deepest near the stripe ends and bounded by steep barriers just beyond them. Because the field beneath the stripe is also lower than in the surrounding film along $x$, this well confines spin waves in both in-plane directions. In the AP configuration the profile is inverted (Fig. \ref{effective-field-comparison}(b)). The field beneath the stripe is raised to about 15 mT at the center and reaches maxima of nearly 50 mT just inside the ends, while shallow minima of about $-5$ mT form just outside them. Along the stripe, the effective field therefore has a minimum at the stripe center, bounded by barriers near the ends, but the region beneath the stripe as a whole lies at a higher field than the surrounding film. Near the slanted ends the stray field also has sizable $x$ and $z$ components, which tilt the YIG magnetization out of the film plane (Fig. \ref{Fig1-crossection}(d)) and break the mirror symmetry discussed in Section \ref{subsec:length-short}.

The field at the stripe center depends strongly on the stripe length (Fig. \ref{effective-field-comparison}(c),(d)). Reducing $L$ from 1.78 µm to 580 nm lowers it from about 9 to 4.5 mT in the P configuration and raises it from about 11 to 15.5 mT in the AP configuration, whereas the $x$ and $z$ components at the center remain negligible. For $L=700$ nm, the deviation from the bias field amounts to about 3.5 mT in the P configuration and 4.5 mT in the AP configuration. The point-charge estimate reproduces the rapid increase of the stray field with decreasing $L$ but overestimates its magnitude by about a factor of two, because the end charges are distributed over the end faces and the YIG film extends over a finite depth below the stripe. The dependence of the stray field on stripe width and thickness (Fig. \ref{effective-field-comparison}(e),(f)) is discussed in Section \ref{sec:width-height}.

These field landscapes account for the different behavior of the two configurations. In the P configuration, the reduced field beneath the stripe lowers the local spin-wave frequency and forms a two-dimensional spin-wave well, analogous to the wells formed by inhomogeneous internal fields in confined magnetic elements \cite{Kostylev2007,Jorzick2002} and to spin-wave channels defined by local field reductions \cite{Qin2022}. The well supports modes quantized along $y$, whose frequencies increase as the well shortens, as observed in Fig. \ref{all-lengths-spectra-comparison-par}(a). In the AP configuration, the stray field raises the mean effective field beneath the stripe, which shifts the resonance to higher frequency as $L$ decreases (Fig. \ref{all-lengths-spectra-comparison-par}(b)). The field maxima near the stripe ends act as barriers for BV spin waves propagating along the stripe and reduce the leakage of spin-wave energy along $y$, so that the precession is confined into a single mode centered in the resonator (Fig. \ref{Fig4_freq_comparison_600nm}(k)), whose enhancement increases as $L$ decreases.

\begin{figure}[h]
	\centering
	\includegraphics[width=0.6\columnwidth]{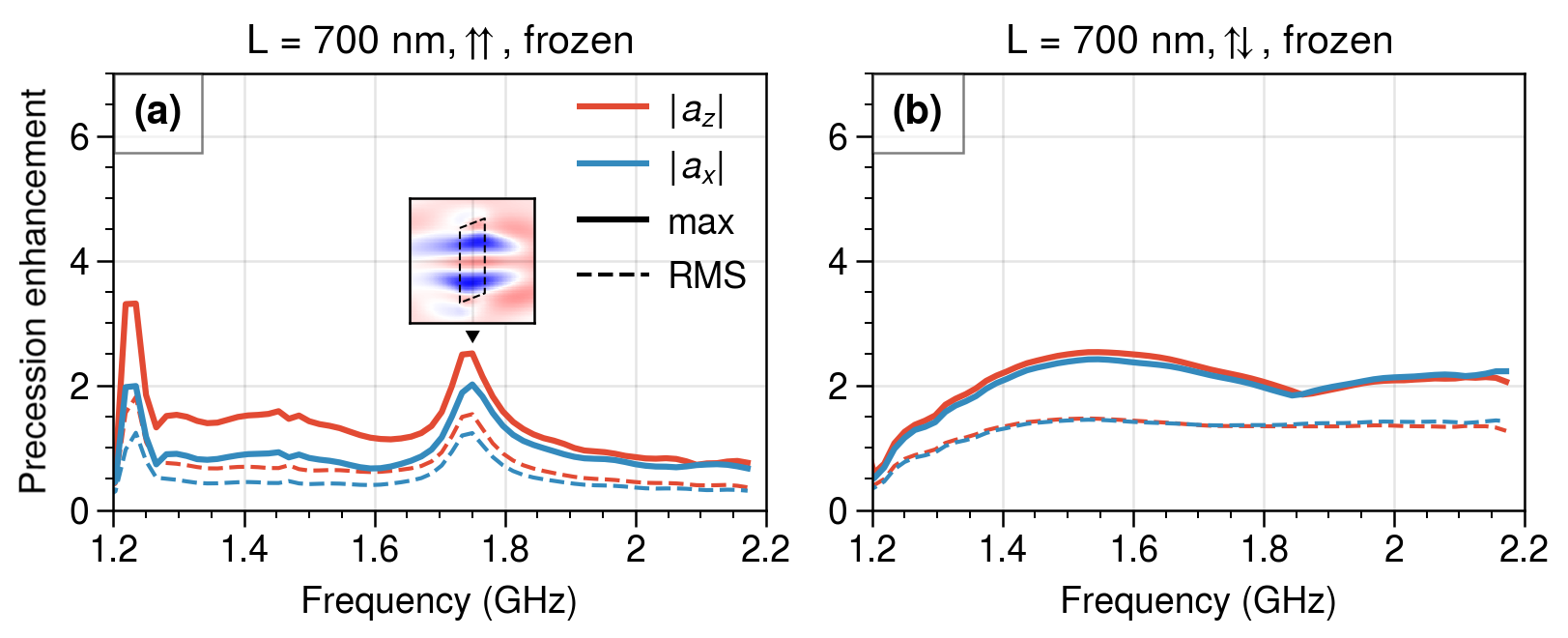}
	\caption{Precession enhancement with ``frozen'' CoFeB magnetization. Maximum (solid lines) and root-mean-square (dashed lines) values of the normalized amplitudes $|a_z|$ (red) and $|a_x|$ (blue) in the top YIG layer beneath a 700-nm-long CoFeB stripe whose magnetization is frozen after relaxation, for the P (a) and AP (b) configurations. Freezing the CoFeB magnetization removes the dynamic dipolar coupling but preserves the static stray field. In the P configuration the quantized modes at about 1.25 and 1.73 GHz persist, whereas in the AP configuration the strong resonance at 1.61 GHz (Fig. \ref{Fig5_three-spectra-comparison}(f)) disappears. The inset in (a) shows the precession profile of the $m=3$ mode at 1.73 GHz.}
	\label{frozen-spins}
\end{figure}

To separate the contributions of the static stray field and the dynamic dipolar coupling, we repeated the simulations for $L=700$ nm with the magnetization dynamics in CoFeB disabled (``frozen'') after relaxation, using the \texttt{FrozenSpins} option of MuMax3. In these simulations the stripe retains its static stray field but its magnetization cannot precess, so that the dynamic dipolar coupling to the YIG film, and with it the Fabry–Pérot mechanism, is switched off. In the P configuration the two sharp resonances persist (Fig. \ref{frozen-spins}(a)), at about 1.25 and 1.73 GHz, close to the resonances at 1.27 and 1.70 GHz obtained with dynamic coupling (Fig. \ref{Fig5_three-spectra-comparison}(c)), and the precession profile identifies the upper resonance as the $m=3$ quantized mode (inset of Fig. \ref{frozen-spins}(a)). In contrast, the broad enhancement between 1.3 and 1.6 GHz, including the precession localized near the stripe ends at 1.39 GHz, disappears. In the AP configuration, freezing the CoFeB magnetization removes the strong resonance at 1.61 GHz entirely, and only a weak, nearly featureless enhancement remains (Fig. \ref{frozen-spins}(b)). Because the region beneath the stripe lies at a higher field than the surrounding film, the static stray field alone cannot confine spin waves along $x$. The strongly enhanced AP mode therefore originates from the Fabry–Pérot cavity formed by the dynamic dipolar coupling between the CoFeB stripe and the YIG film, which confines the precession along $x$, while the static stray field confines it along $y$ and sets its frequency.

The static stray field thus constitutes a second control parameter for short resonators, alongside the dynamic dipolar coupling that defines the Fabry–Pérot cavity. In the P configuration it dominates the response and turns the resonator into a field-defined spin-wave well with several weakly enhanced modes, whereas in the AP configuration it cooperates with the dynamic coupling to produce a single strongly enhanced mode. Because reversing the stripe magnetization inverts the stray-field landscape as well as the chirality of the dynamic coupling, switching between the two configurations reconfigures the resonator in a non-volatile manner. 

\begin{figure}[b!]
	\centering
	\includegraphics[width=\columnwidth]{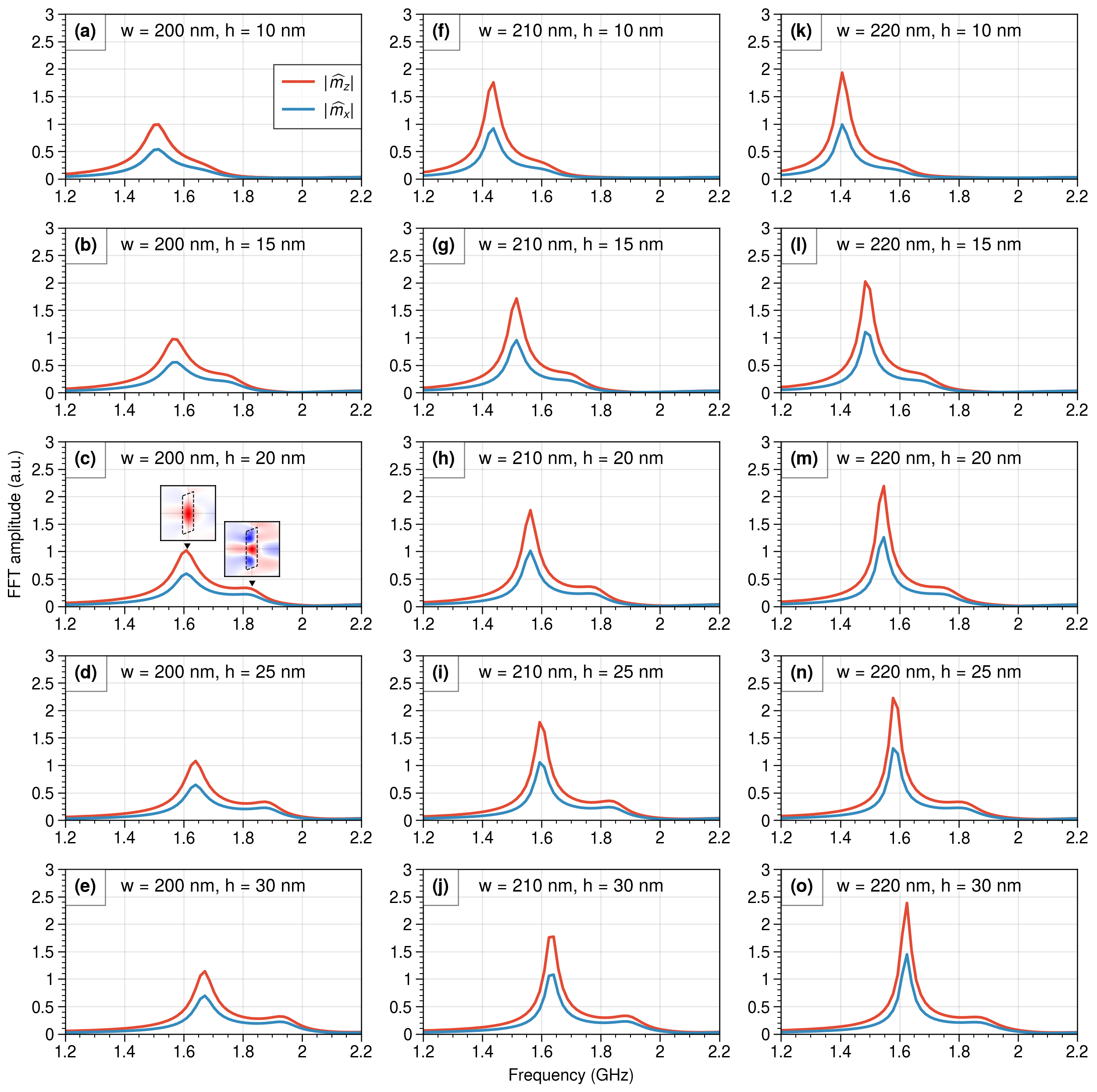}
	\caption{Width and thickness tuning of the AP nanoresonator. Precession amplitude spectra ($|\yhwidehat{m}_x|$ blue; $|\yhwidehat{m}_z|$ red) extracted from the top YIG layer beneath the center of a 700-nm-long stripe, driven by a uniform in-plane field along $x$ confined to the CoFeB. The stripe Width $w$ increases from 200 to 220 nm (left to right) and the stripe thickness $h$ increases from 10 to 30 nm (top to bottom). Insets in (c) show the $m=1$ and $m=3$ mode profiles. Amplitudes are FFT magnitudes in arbitrary units and are not normalized to the bare YIG film.}
	\label{varying-widths-heights-comparison}
\end{figure}

\section{\label{sec:width-height}Tuning the AP nanoresonator: width and thickness}
The resonance frequency and enhancement of the AP nanoresonator can be tuned by small changes of its width and thickness. Figure \ref{varying-widths-heights-comparison} shows precession spectra of a 700-nm-long AP resonator for $w=200-220$ nm and $h=10-30$ nm, excited by a uniform in-plane field confined to the CoFeB stripe (Section \ref{sec:methods}). This excitation isolates the intrinsic resonator response from its coupling to incident spin waves. For the reference geometry ($w=200$ nm, $h=20$ nm, Fig. \ref{varying-widths-heights-comparison}(c)), the main resonance appears at 1.61 GHz, as in Fig. \ref{Fig5_three-spectra-comparison}(f), together with a weak second resonance $0.15-0.20$ GHz higher. The precession profiles identify these as the first ($m=1$) and third ($m=3$) quantized modes along $y$. Modes of even order are not excited, because a uniform field has no overlap with precession profiles that are antisymmetric along $y$.

Width and thickness shift the resonance in opposite directions, because they act predominantly through different mechanisms. 
Increasing the stripe width by 5–10\% (to 210 and 220 nm) lowers the resonance frequency by 50–100 MHz and increases the resonance amplitude by up to 100\% (compare Fig. \ref{varying-widths-heights-comparison}(c) and \ref{varying-widths-heights-comparison}(m)). The accompanying reduction of the linewidth suggests weaker radiative loss of the wider cavity. The frequency decrease follows from the resonance condition ($|k_2| + |k_3|)w + \phi_0 = (2n + 1)\pi$ \cite{Qin2021}, which a wider cavity satisfies at smaller wavenumbers and thus at a lower frequency on the bilayer dispersion. A wider stripe also produces a larger static stray field, because the magnetic charge on its ends scales with the stripe cross-section $wh$. In the AP configuration, the effective field in YIG at the stripe center increases from about 13.5 to 15 mT when $w$ is increased from 180 to 240 nm (Fig. \ref{effective-field-comparison}(e)), which by itself would raise the resonance frequency. Over the width range of Fig. \ref{varying-widths-heights-comparison}, however, this increase amounts to only about 0.4 mT, and the downshift imposed by the resonance condition dominates. 

Increasing the stripe thickness by 25–50\% (to 25 and 30 nm) raises the resonance frequency by a similar 50–100 MHz, with a weaker effect on the amplitude (compare Fig. \ref{varying-widths-heights-comparison}(c) and \ref{varying-widths-heights-comparison}(e)). A thicker stripe strengthens the dynamic dipolar coupling, which by itself would lower the frequency \cite{Qin2021}. At the same time, the effective field in YIG at the stripe center increases from about 12 to 16 mT when $h$ is increased from 10 to 30 nm (Fig. \ref{effective-field-comparison}(f)). Between 20 and 30 nm, this corresponds to an increase of about 2 mT, five times larger than the change produced by the width variation in Fig. \ref{varying-widths-heights-comparison}. The stray field therefore outweighs the stronger dynamic coupling, and the resonance frequency increases with thickness. On average, the resonance frequency thus changes by about $-5$ MHz per nanometer of width and $+10$ MHz per nanometer of thickness. Because the two parameters shift the AP resonance in opposite directions, their combination allows a high-quality resonance to be placed at a target frequency, which is essential for networks in which nodes must be matched or deliberately detuned.

\begin{figure}[h]
	\centering
	\includegraphics[width=\columnwidth]{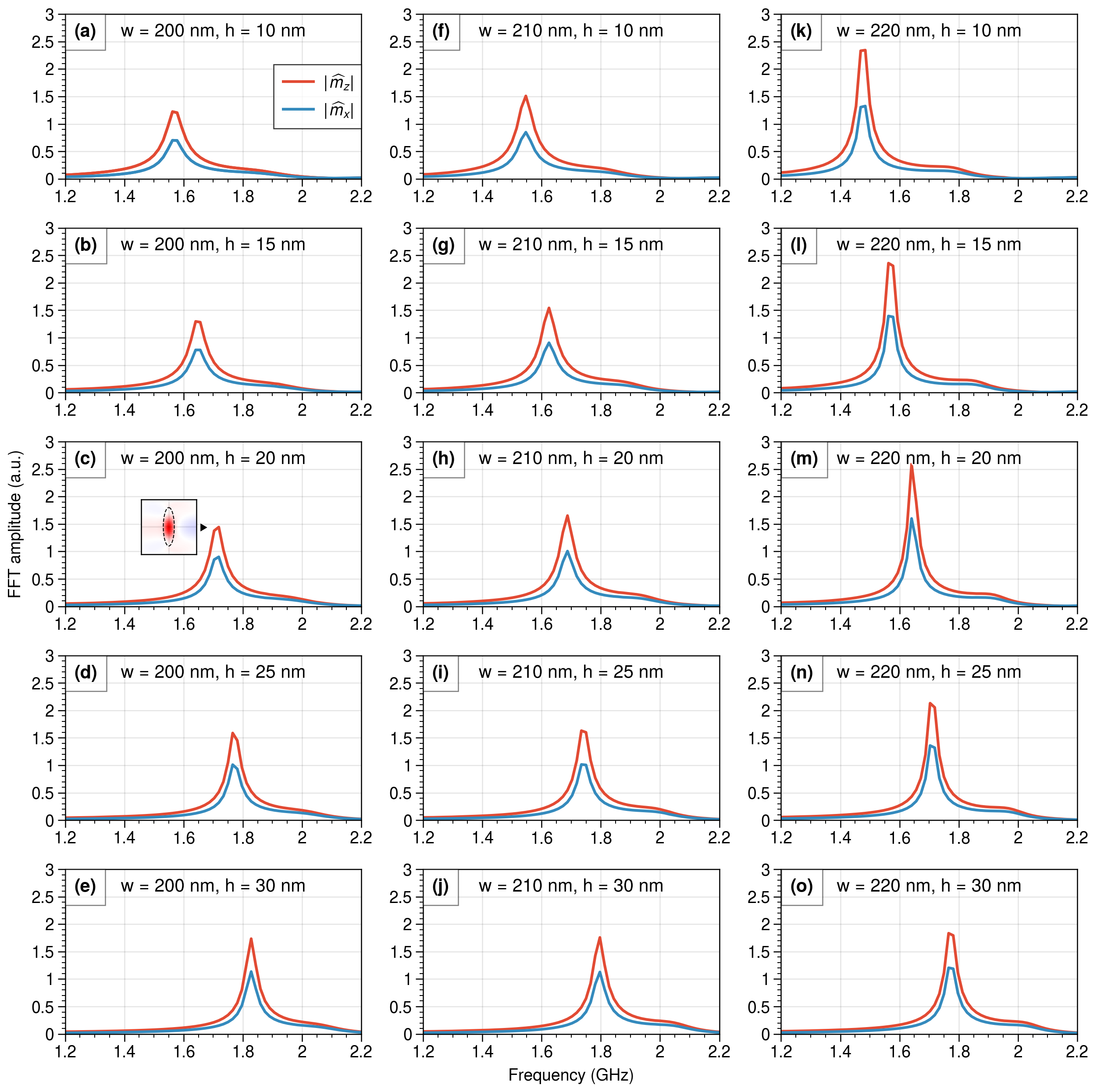}
	\caption{Elliptical resonators. As Fig. \ref{varying-widths-heights-comparison}, for elliptical stripes with major axis $L=700$ nm along $y$ and minor axis $w$. The inset in (c) shows the profile of the main $m=1$ mode. The $m=3$ mode is suppressed in elliptical resonators.}
	\label{ellipse-widths-heights-comparison}
\end{figure}

\section{\label{sec:ellipse}Robustness to resonator shape}
In electron-beam lithography, the finite beam size rounds the corners of rectangular patterns, and elliptical shapes require fewer exposure steps. To assess the robustness of the AP resonance to the stripe shape, we replaced the slanted rectangle by an ellipse with major axis $L=700$ nm along $y$ and minor axis $w$ along $x$ (Fig. \ref{ellipse-widths-heights-comparison}). The sharp main resonance is preserved for all widths and thicknesses studied, whereas the $m=3$ mode is suppressed compared with the rectangular stripe. This is consistent with the more uniform effective field of elliptical elements \cite{Osborn1945}, which lack the regions of strongly nonuniform field near rectangular ends (Fig. \ref{effective-field-comparison}). The main resonance of the elliptical resonator lies 0.1–0.2 GHz above that of a rectangular resonator with the same nominal width and thickness. Part of this shift reflects the smaller mean width of the ellipse, which raises the frequency through the resonance condition discussed above. The resonance amplitude is preserved and in some geometries exceeds that of the corresponding rectangular resonator by up to 20\%. Elliptical AP resonators therefore combine a cleaner single-mode spectrum with strong precession enhancement and simpler fabrication.

\section{\label{sec:nonlin}Nonlinear response}
The strong precession enhancement of the AP nanoresonator drives the YIG beneath the stripe into the nonlinear regime at incident spin-wave amplitudes of only a few percent of the saturation magnetization. Resonant energy concentration is known to lower nonlinear thresholds in ring, chiral and Fabry–Pérot resonators \cite{Wang2020,Fripp2023,Lutsenko2026}. For a 15-µm-long Fabry–Pérot resonator, the nonlinear response was characterized by the transmission coefficient, which exhibits a power-dependent downshift of the transmission gaps \cite{Lutsenko2026}. For a sub-micrometer resonator, scattering into all directions makes the transmitted amplitude dependent on the observation point. We therefore evaluate the nonlinear response by the precession amplitudes $|m_x| = |M_x|/M_s$ and $|m_z| = |M_z|/M_s$ at the resonator center.

\begin{figure}[ht!]
	\centering
	\includegraphics[width=0.85\columnwidth]{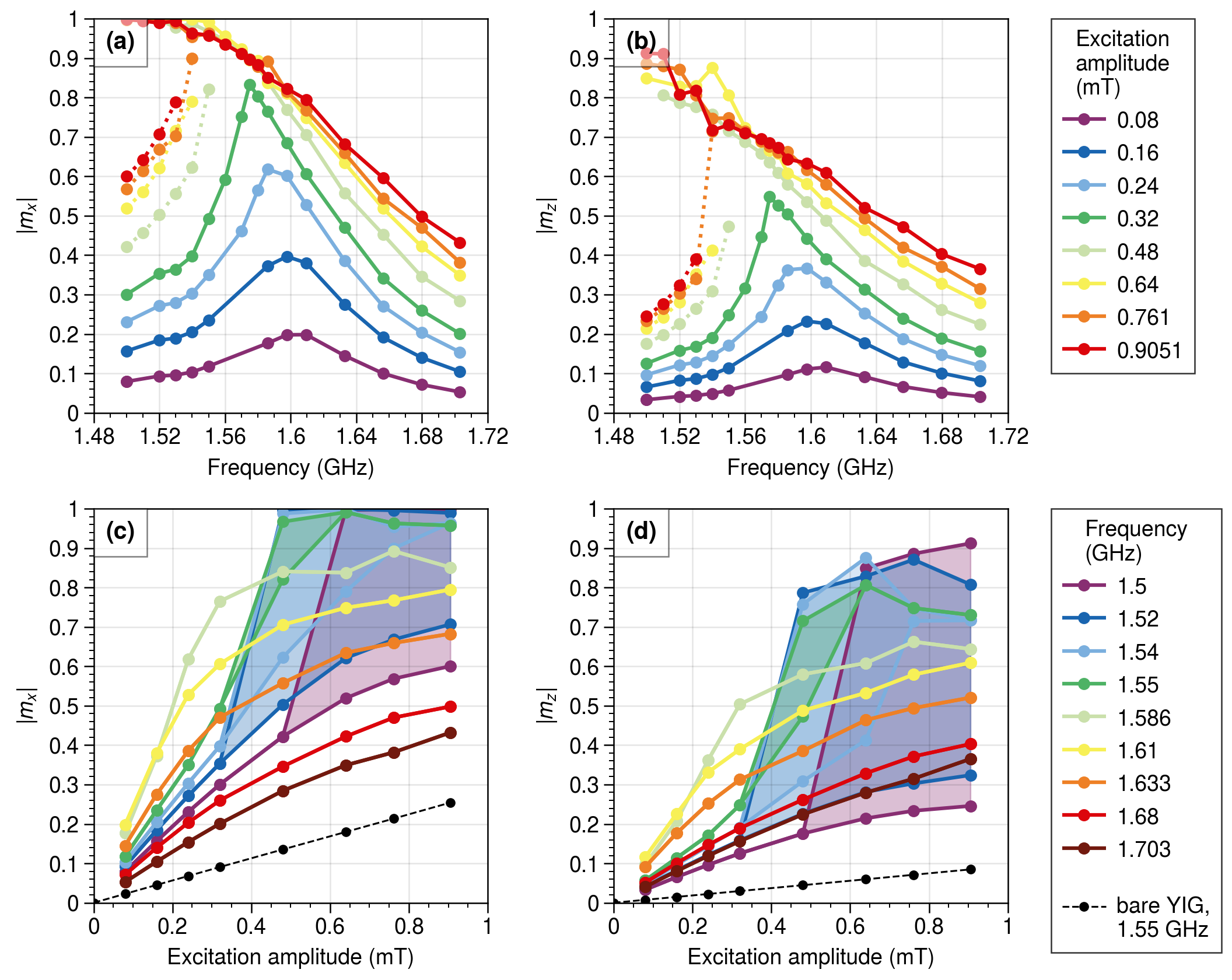}
	\caption{Nonlinear response of the 700-nm AP resonator. (a) In-plane and (b) out-of-plane precession amplitudes in the top YIG layer beneath the center of the stripe versus frequency, for excitation fields of 0.08–0.91 mT. Dotted segments mark the lower branches in the bistable region; fold-over occurs below approximately 1.55 GHz at the highest excitation field. (c) In-plane and (d) out-of-plane resonator precession amplitudes versus excitation field at selected frequencies. The black lines show the precession amplitudes in the bare YIG film at 1.55 GHz. Departure from linearity in the resonator sets in at incident amplitudes as small as $|m_x|\approx0.05$ and $|m_z|\approx0.015$.}
	\label{nonlinear}
\end{figure}

Figure \ref{nonlinear} shows the precession amplitudes in the top YIG layer at the center of a 700-nm-long AP resonator for excitation fields between 0.08 and 0.91 mT, which correspond to incident spin-wave amplitudes in bare YIG of $|m_x|=0.023-0.26$ and $|m_z|=0.006-0.08$. Figure \ref{nonlinear}(a),(b) shows resonance curves of the in-plane and out-of-plane components for each excitation field, and Fig. \ref{nonlinear}(c),(d) shows both components as a function of excitation field at fixed frequencies. At the lowest excitation field of 0.08 mT, the resonance curve is symmetric and centered close to 1.60 GHz, consistent with the resonance at 1.61 GHz obtained under broadband excitation, and its full width at half maximum is about 100 MHz. The resonator precession reaches $|m_x|\approx0.20$ and $|m_z|\approx0.11$, which is about 9 and 17 times the corresponding amplitudes in bare YIG. These ratios agree reasonably with the in-plane and out-of-plane enhancements of Fig. \ref{Fig5_three-spectra-comparison}(f) and confirm that the response at 0.08 mT is linear. The ratio $|m_z|/|m_x|\approx0.6$ within the resonator is about twice that of the incident spin wave in bare YIG, showing that the dipolar coupling to the stripe makes the precession considerably less elliptical.

Doubling the excitation field to 0.16 mT, which corresponds to an incident $|m_x|\approx0.05$ and $|m_z|\approx0.015$, already shifts the resonance to lower frequency and makes the curve slightly asymmetric, with a steeper low-frequency flank. Shift and asymmetry are identical for both components (Fig. \ref{nonlinear}(a),(b)), confirming that they reflect the same resonant mode. At this field the precession amplitudes within the resonator are $|m_x|\approx0.40$ and $|m_z|\approx0.23$, corresponding to a precession cone of more than 20$^\circ$, whereas the bare YIG film responds linearly in both components over the entire field range (black lines in Fig. \ref{nonlinear}(c),(d)). With further increase of the excitation field, the resonance shifts down by about 15 MHz at 0.24 mT and about 30 MHz at 0.32 mT, where $|m_x|\approx0.84$ and $|m_z|\approx0.55$, and the low-frequency flank becomes nearly vertical in both components. This line shape is characteristic of a Duffing-type oscillator with a softening nonlinearity that approaches its critical point, at which the nonlinear frequency shift becomes comparable to the half-width of the resonance \cite{Anderson1955,Fetisov1999,Gui2009}. 

At 0.48 mT, corresponding to an incident $|m_x|\approx0.14$ and $|m_z|\approx0.04$, the critical point is exceeded and the resonance folds over toward lower frequency. A low-amplitude branch (dotted lines) and a high-amplitude branch (solid lines) then coexist over a finite frequency range, and the state that is reached depends on the excitation history. This bistable range widens with increasing excitation field and, at 0.91 mT, extends from below 1.49 GHz, the lowest frequency simulated, to about 1.53 GHz. On the high-amplitude branch the resonance has shifted down by more than 100 MHz, which exceeds the linear linewidth. The two components distinguish the branches with different contrast. Because $|m_x|$ already approaches 0.8 on the low-amplitude branch and saturates near 1 on the high-amplitude branch, the in-plane contrast between the branches is small. The out-of-plane component, in contrast, approximately doubles between the branches, from about 0.4 to 0.9 at 0.91 mT. The negative sign of the nonlinear frequency shift agrees with theoretical predictions for in-plane magnetized films in which the spin dynamics is dominated by dipolar interactions \cite{Krivosik2010} and with the experimentally observed downshift of Fabry–Pérot transmission gaps at high excitation power \cite{Lutsenko2026}. The short wavelength of the $λ_3$ mode in the bilayer is expected to strengthen the nonlinearity further, because the nonlinear frequency shift of dipole-exchange spin waves increases with wavenumber \cite{Krivosik2010}.

On the high-amplitude branch, $|m_x|$ saturates near 1 (Fig. \ref{nonlinear}(a)), which implies a precession cone approaching 90$^\circ$. The out-of-plane amplitude does not saturate in the same manner but continues to grow with excitation field and reaches about 0.9 at 0.91 mT (Fig. \ref{nonlinear}(b)). The ratio $|m_z|/|m_x|$ therefore increases from about 0.6 in the linear regime to about 0.9, and the precession becomes nearly circular at large angles. This is consistent with the reduction of the static magnetization component $M_y$ at large precession angles, which weakens the dynamic demagnetizing field responsible for the ellipticity of small-angle precession in thin films \cite{Kalinikos1986}. At such amplitudes the precession is no longer harmonic, and the Fourier amplitude at the excitation frequency captures only part of the magnetization dynamics.

The dependence on excitation field at fixed frequency shows that a single resonator provides distinct nonlinear transfer functions, depending on the position of the operating frequency relative to the linear resonance. Because the resonance shifts to lower frequency with increasing precession amplitude, the nonlinearity acts in opposite directions on the two sides of the linear resonance. Below the linear resonance frequency, the shifting resonance moves toward the excitation frequency. The precession amplitude therefore first grows slowly and then jumps to the high-amplitude branch once the fold-over threshold is exceeded, as seen in Fig. \ref{nonlinear}(a),(b) between 1.50 and 1.55 GHz. This produces a threshold-like activation with hysteresis (Fig. \ref{nonlinear}(c),(d)), whose step is most pronounced in $|m_z|$. The hysteresis within the bistable range further provides a form of short-term memory, which is of interest for recurrent and reservoir computing schemes \cite{Watt2021,Korber2023,Leenders2026}.

Slightly below the linear resonance, the shifting resonance reaches the excitation frequency already at moderate amplitudes and then moves past it. At 1.586 GHz, $|m_x|$ increases from 0.18 to 0.77 between 0.08 and 0.32 mT, initially slightly faster than linearly, and then gradually saturates near 0.85, although the excitation field increases almost threefold further (Fig. \ref{nonlinear}(c)). The out-of-plane component follows the same trend (Fig. \ref{nonlinear}(d)). It increases from 0.10 to 0.50 over the same initial field range and grows by a further 25\% to 0.64 at 0.91 mT. The enhancement relative to bare YIG decreases accordingly, from about 8 to 3.5 for the in-plane component and from about 17 to 8 for the out-of-plane component. This combination of a steep initial rise and saturation gives a sigmoid-like transfer function. 

Above the linear resonance frequency, the resonance moves away from the excitation frequency from the outset, so that the excitation falls increasingly onto the high-frequency tail of the line and the response is sublinear at all fields. Just above the resonance, the precession amplitude is large but deviates most strongly from proportionality. At 1.633 GHz, $|m_x|$ increases from about 0.15 to 0.68 while the excitation field increases more than tenfold, so that the in-plane enhancement relative to bare YIG drops from about 6.5 to 2.6 and the resonator acts as a saturating, limiter-like element. With increasing distance from the resonance the nonlinear shift has less effect on the response, and hence, the deviation from proportionality decreases in both components. At 1.703 GHz the precession amplitude is smaller, but $|m_x|$ and $|m_z|$ grow nearly in proportion to the excitation field, and the in-plane enhancement changes only from about 2.2 to 1.7, which corresponds to an almost linear response. 

Threshold activation and nonlinear suppression have been observed in transmission experiments on long Fabry–Pérot resonators, where they were described as ReLU-like activation and nonlinear suppression \cite{Lutsenko2026}. Here these responses occur in a resonator with a footprint of $200\times700$ nm$^2$, and the choice of operating frequency selects between them. 

In summary, the response of the resonator becomes nonlinear at an excitation field of 0.16 mT, which corresponds to incident spin-wave amplitudes in the bare YIG film of $|m_x|\approx0.05$ and $|m_z|\approx0.015$. Fold-over sets in at 0.48 mT, corresponding to $|m_x|\approx0.14$ and $|m_z|\approx0.04$. The bare YIG film, in contrast, responds linearly over the entire range of excitation fields studied. The 700-nm-long AP resonator thus provides a strong nonlinearity at spin-wave amplitudes readily accessible with on-chip antennas. Its footprint is smaller than that of nonlinear ring resonators \cite{Wang2020} and its length is more than an order of magnitude shorter than that of previously studied Fabry–Pérot resonators \cite{Lutsenko2026}.

\section{\label{sec:conclusions}Discussion and conclusions}
Our simulations show how a magnonic Fabry–Pérot resonator evolves from a one-dimensional cavity into a sub-micrometer two-dimensional scatterer as the length of its CoFeB stripe is reduced. An infinitely long  and 200-nm-wide stripe on a YIG film behaves as an ideal Fabry–Pérot cavity. It opens a narrow transmission gap at the $n=1$ resonance, leaves the transmitted wavefront flat and enhances the out-of-plane precession in the YIG beneath the stripe by factors of about 5 and 12 in the P and AP configurations, respectively. For stripes a few micrometers long, the Fabry–Pérot resonance persists at nearly the same frequency, but BV spin waves emitted from the stripe ends interfere with the resonant precession. The resulting modulation along the stripe follows the BV dispersion of bare YIG, because the chiral dipolar coupling to the stripe vanishes for wave vectors parallel to the magnetization, and the reflected and transmitted wavefronts are fragmented. Below about 1.2 µm, the two magnetic configurations diverge. In the P configuration the Fabry–Pérot resonance breaks up into several quantized modes with reduced enhancement, whereas in the AP configuration a single mode survives, shifts to higher frequency and reaches an out-of-plane enhancement of about 15, exceeding that of the infinite resonator. Simulations of the static stray field and of resonators with frozen CoFeB magnetization show that this dichotomy arises from the interplay between the static stray field of the stripe ends and the dynamic dipolar coupling. In the P configuration the stray field forms a spin-wave well that supports quantized modes even without dynamic coupling, whereas in the AP configuration it confines the dynamically coupled Fabry–Pérot precession along the stripe into a single strongly enhanced mode. At all finite lengths, and most prominently below 1.2 µm, the resonator scatters the incident plane wave over the entire isofrequency contour of the YIG film, including caustic beams, with an angular distribution determined by the chirality of the dipolar coupling and the geometry of the stripe ends.

The AP nanoresonator meets the requirements for a node in a magnonic neural network. Its footprint of $200\times700$ nm$^2$ is far below the incident spin-wave wavelength in both in-plane directions, and it scatters spin waves over a wide angular range, so that a single node can couple to many others. Its strong precession enhancement lowers the onset of nonlinearity to incident amplitudes of about $|m_x|\approx0.05$ and $|m_z|\approx0.015$, while fold-over occurs at $|m_x|\approx0.14$ and $|m_z|\approx0.04$. Depending on whether the excitation lies above or below the linear resonance, the same resonator emulates threshold-like activation, a sigmoid-like transfer function, or a saturating, limiter-like element. Switching the resonator between the P and AP configurations changes its resonance frequency, its magnetization precession enhancement and its scattering pattern in a non-volatile manner and thus provides a programmable weight \cite{Qin2021, Lutsenko2025}. The resonance frequency can be adjusted within about 100 MHz by nanometer-scale changes of width and thickness, and an elliptical stripe yields a cleaner single-mode response, while also being easier to fabricate. In contrast to chiral magnonic resonators, in which the resonant precession and the nonlinearity reside in the metallic element \cite{Fripp2023}, the resonant energy here is stored in the low-damping YIG film.

These properties suggest two further directions. Because each resonator couples efficiently to a uniform microwave field, a global field could supply energy locally to every node of a network without dedicated waveguides \cite{Kruglyak2021}. Moreover, the dependence of resonance frequency, enhancement and scattering pattern on the position, size and shape of each resonator provides natural parameters for inverse design, including the design of deterministic nonlinear functions \cite{Wang2021,Vilsmeier2026}. Experimental verification is within reach, since the incident and scattered spin-wave wavelengths can be resolved by SNS-MOKE microscopy \cite{Qin2021,Lutsenko2025,Dreyer2021} and the required stripes are well within the capabilities of standard electron-beam lithography.

\ack{
This project has received funding from the European Union's Horizon Europe research and innovation program under Grant Agreement No. 101070347-MANNGA. However, views and opinions expressed are those of the authors only and do not necessarily reflect those of the European Union or the European Health and Digital Executive Agency (HADEA). Neither the European Union nor the granting authority can be held responsible for them. We acknowledge the computational resources provided by the Aalto Science-IT project.}

\bibliographystyle{iopart-num-doi}
\bibliography{references-new}% Produces the bibliography via BibTeX.

\end{document}